\documentclass[12pt]{article}
\usepackage{axodraw,bbold}

\catcode`@=12
\begin{document}
\thispagestyle{empty}
\begin{flushright} 
UCRHEP-T416\\ 
July 2006\
\end{flushright}
\vspace{0.5in}
\begin{center}
{\LARGE	\bf Neutrino Mass Matrix from $\Delta(27)$ Symmetry\\}
\vspace{1.5in}
{\bf Ernest Ma\\}
\vspace{0.2in}
{\sl Physics Department, University of California, Riverside, 
California 92521 \\}
\vspace{1.5in}
\end{center}

\begin{abstract}\
The discrete subgroup $\Delta(27)$ of $SU(3)$ has some interesting 
properties which may be useful for understanding charged-lepton and 
neutrino mass matrices.  Assigning leptons to the {\bf 3} and 
$\bar{\bf 3}$ representations of $\Delta(27)$, a simple form of the 
Majorana neutrino mass matrix is obtained and compared to present data.
\end{abstract}

\newpage
\baselineskip 24pt

Since the introduction of the discrete symmetry $A_4$ \cite{mr01,bmv03} for 
understanding the family structure of leptons, much progress has been made 
\cite{m04,af05,m05-1,bh05,m06,af06,mvkr05} in obtaining the so-called 
tribimaximal mixing pattern of Harrison, Perkins, and Scott 
\cite{hps02,hz03}.  Whereas $A_4$ is the group of even permutation of four 
objects, it is also the symmetry group of the perfect tetrahedron 
\cite{plato}, and identical to the subgroup $\Delta(12)$ of $SU(3)$.  
The next subgroup in the series $\Delta(3n^2)$ is $\Delta(27)$ \cite{su3}, 
which was considered in a model of T violation \cite{bgg84} and has just 
recently been applied \cite{mvkr06} to quark and lepton mass matrices.  
In this note, a minimal alternative for leptons is proposed, which results 
in a Majorana neutrino mass matrix of the form
\begin{equation}
{\cal M}_\nu = \pmatrix{fa & c & b \cr c & fb & a \cr b & a & fc},
\end{equation}
in the basis where ${\cal M}_l$ is diagonal.  For comparison, two 
previously proposed models based on $A_4$ have \cite{hmvv05}
\begin{equation}
{\cal M}_\nu = \pmatrix{a & d & d \cr d & b & d \cr d & d & c},
\end{equation}
and \cite{m05-2}
\begin{equation}
{\cal M}_\nu = \pmatrix{fa & \sqrt{ab} & \sqrt{ac} \cr \sqrt{ab} & fb & 
\sqrt{bc} \cr \sqrt{ac} & \sqrt{bc} & fc},
\end{equation}
respectively.  The mass matrices of Eqs.~(1) and (2) have 4 independent 
moduli and 3 independent phases, having thus 2 predictions, whereas that 
of Eq.~(3) has 4 independent moduli and only 1 independent phase, having 
thus 4 predictions.  In the limit $b=c$ \cite{m02}, they all have 3 
independent moduli and 1 independent phase, so that there are five 
predictions, three of which are common to all three models, i.e. 
$\theta_{23} = \pi/2$, $\theta_{13}=0$, and the CP nonconserving Dirac 
phase is irrelevant (because $\theta_{13}=0$).

The non-Abelian discrete group $\Delta(27)$ has 27 elements divided into 11 
equivalence classes.  It has 9 one-dimensional irreducible representations 
${\bf 1_i} (i=1,...,9$) and 2 three-dimensional ones {\bf 3} and 
$\bar{\bf 3}$.  Its character table is given below, where $n$ is the number 
of elements, $h$ is the order of each element, and $\omega = \exp(2 \pi i/3)$ 
with $1+\omega+\omega^2=0$.

\begin{table}[htb]
\caption{Character table of $\Delta(27)$.}
\begin{center}
\begin{tabular}{|c|c|c|c|c|c|c|c|c|c|c|c|c|c|}
\hline 
Class & $n$ & $h$ & ${\bf 1_1}$ & ${\bf 1_2}$ & ${\bf 1_3}$ & ${\bf 1_4}$ 
& ${\bf 1_5}$ & ${\bf 1_6}$ & ${\bf 1_7}$ & ${\bf 1_8}$ & ${\bf 1_9}$ 
& ${\bf 3}$ & $\bar{\bf 3}$ \\ 
\hline
$C_1$ & 1 & 1 & 1 & 1 & 1 & 1 & 1 & 1 & 1 & 1 & 1 & 3 & 3 \\ 
$C_2$ & 1 & 3 & 1 & 1 & 1 & 1 & 1 & 1 & 1 & 1 & 1 & 3$\omega$ & 3$\omega^2$ \\ 
$C_3$ & 1 & 3 & 1 & 1 & 1 & 1 & 1 & 1 & 1 & 1 & 1 & 3$\omega^2$ & 3$\omega$ \\ 
$C_4$ & 3 & 3 & 1 & $\omega$ & $\omega^2$ & 1 & $\omega^2$ & $\omega$ & 1 & 
$\omega$ & $\omega^2$ & 0 & 0 \\ 
$C_5$ & 3 & 3 & 1 & $\omega^2$ & $\omega$ & 1 & $\omega$ & $\omega^2$ & 1 & 
$\omega^2$ & $\omega$ & 0 & 0 \\ 
$C_6$ & 3 & 3 & 1 & 1 & 1 & $\omega^2$ & $\omega^2$ & $\omega^2$ & $\omega$ & 
$\omega$ & $\omega$ & 0 & 0 \\ 
$C_7$ & 3 & 3 & 1 & $\omega$ & $\omega^2$ & $\omega^2$ & $\omega$ & 1 & 
$\omega$ & $\omega^2$ & 1 & 0 & 0 \\ 
$C_8$ & 3 & 3 & 1 & $\omega^2$ & $\omega$ & $\omega^2$ & 1 & $\omega$ & 
$\omega$ & 1 & $\omega^2$ & 0 & 0 \\ 
$C_9$ & 3 & 3 & 1 & 1 & 1 & $\omega$ & $\omega$ & $\omega$ & $\omega^2$ & 
$\omega^2$ & $\omega^2$ & 0 & 0 \\ 
$C_{10}$ & 3 & 3 & 1 & $\omega^2$ & $\omega$ & $\omega$ & $\omega^2$ & 1 
& $\omega^2$ & $\omega$ & 1 & 0 & 0 \\ 
$C_{11}$ & 3 & 3 & 1 & $\omega$ & $\omega^2$ & $\omega$ & 1 & $\omega^2$ 
& $\omega^2$ & 1 & $\omega$ & 0 & 0 \\ 
\hline
\end{tabular}
\end{center}
\end{table}

The group multiplication rules are
\begin{equation}
{\bf 3} \times {\bf 3} = \bar{\bf 3} + \bar{\bf 3} + \bar{\bf 3},
~~{\rm and}~~{\bf 3} \times \bar{\bf 3} = \sum^9_{i=1} {\bf 1_i},
\end{equation}
where
\begin{eqnarray}
&& {\bf 1_1} = 1 \bar 1 + 2 \bar 2 + 3 \bar 3, ~~~{\bf 1_2} = 1 \bar 1 + 
\omega 2 \bar 2 + \omega^2 3 \bar 3, ~~~{\bf 1_3} = 1 \bar 1 + \omega^2 2 
\bar 2 + \omega 3 \bar 3, \\
&& {\bf 1_4} = 1 \bar 2 + 2 \bar 3 + 3 \bar 1, ~~~{\bf 1_5} = 1 \bar 2 + 
\omega 2 \bar 3 + \omega^2 3 \bar 1, ~~~{\bf 1_6} = 1 \bar 2 + \omega^2 2 
\bar 3 + \omega 3 \bar 1, \\
&& {\bf 1_7} = 2 \bar 1 + 3 \bar 2 + 1 \bar 3, ~~~{\bf 1_8} = 2 \bar 1 + 
\omega^2 3 \bar 2 + \omega 1 \bar 3, ~~~{\bf 1_9} = 2 \bar 1 + \omega 3 
\bar 2 + \omega^2 1 \bar 3.
\end{eqnarray}

Let the lepton doublets $(\nu_i,l_i)$ transform as ${\bf 3}$ under 
$\Delta(27)$ and the lepton singlets $l^c_i$ as $\bar{\bf 3}$, then 
with three Higgs doublets transforming as ${\bf 1_1},{\bf 1_2},{\bf 1_3}$, 
the charged-lepton mass matrix is diagonal and has three independent 
masses.  At the same time, with three Higgs triplets transforming as 
${\bf 3}$, the form of Eq.~(1) is obtained.  To see this, consider 
the product ${\bf 3} \times {\bf 3} \times {\bf 3}$.  From Eq.~(4), it is 
clear that it contains three $\Delta(27)$ invariants, i.e. 
$123 + 231 + 312 - 213 - 321 - 132$ [which is invariant under $SU(3)$], 
$123 + 231 + 312 + 213 + 321 + 132$ [which is also invariant under $A_4$], 
and $111 + 222 + 333$.  Since a Majorana mass matrix has to be symmetric, 
only the latter two are allowed.  Without any loss of generality, the 
vacuum expectation values of $\xi^0_{1,2,3}$ may then be taken in the 
proportion $a:b:c$.

Given the form of Eq.~(1), the limit $\theta_{13}=0$ requires $b=c$. 
Under this latter assumption and rotating to the 
basis $[\nu_e, (\nu_\mu + \nu_\tau)/\sqrt 2, (-\nu_\mu + \nu_\tau)
/\sqrt 2]$, Eq.~(1) becomes
\begin{equation}
{\cal M}_\nu = \pmatrix{fa & \sqrt 2 b & 0 \cr \sqrt 2 b & a+fb & 0 
\cr 0 & 0 & fb-a},
\end{equation}
which exhibits maximal $\nu_\mu-\nu_\tau$ mixing, i.e. $\theta_{23}=\pi/4$, 
and $\theta_{13}=0$.  Since $\Delta m^2_{sol} << \Delta m^2_{atm}$ 
experimentally, consider first the limit $\Delta m^2_{sol} = \Delta m^2_{12} 
\to 0$. This has two solutions: either $b \to 0$ and $(f-1)a \to 0$, or 
$fa+a+fb \to 0$.  The former leads to $\Delta m^2_{atm}  
\to 0$, which must be discarded.  Hence
\begin{equation}
a \simeq -fb/(f+1)
\end{equation}
will be assumed from now on.
The mixing angle of the $2 \times 2$ submatrix can be simply read off as
\begin{equation}
\tan 2 \theta_{12} = {2 \sqrt 2 b \over a+fb-fa} \simeq 2 \sqrt 2 \left[ 
1 + {(f-1)(2f+1) \over f+1} \right]^{-1},
\end{equation}
which reduces to $2 \sqrt 2$ in the limit $f=1$ or $f=-1/2$. This would imply 
$\tan^2 \theta_{12} = 1/2$, resulting in tribimaximal mixing \cite{hps02}. 
However Eq.~(9) also implies that
\begin{equation}
\Delta m^2_{atm} \equiv m_3^2 - {m_2^2+m_1^2 \over 2} \simeq 
{2(f-1)(2f+1) b^2 \over f+1},
\end{equation}
which would vanish as well.  Hence this model requires $\tan^2 \theta$ to 
be different from 1/2, and for the experimental value of $0.45 \pm 0.05$, 
it also requires $m_3^2 > m_{1,2}^2$, i.e. a normal ordering of neutrino 
masses.

The effective neutrino mass $m_{ee}$ measured in neutrinoless double beta 
decay is simply given by the magnitude of the $\nu_e \nu_e$ entry of 
${\cal M}_\nu$, i.e. $|fa|$.  For $\tan^2 \theta_{12} = 0.45$, $f=1.1046$ 
or $-0.5248$, resulting in
\begin{equation}
m_{ee} = |fa| = 0.05~{\rm eV}
\end{equation}
in both cases, assuming $\Delta m^2_{atm} = 2.5 \times 10^{-3}$ eV$^2$.  
This value is within reach of the next generation of 
neutrinoless-double-beta-decay experiments.

A variation of this model is to have $l^c_i$ transforming as {\bf 3} 
instead of $\bar{\bf 3}$.  In that case, three Higgs doublets 
transforming as {\bf 3} are required, resulting in
\begin{equation}
{\cal M}_l = \pmatrix{A & f_1 C & f_2 B \cr f_2 C & B & f_1 A \cr f_1 B 
& f_2 A & C}
\end{equation}
For small values of $f_{1,2}$, this reduces to the model being discussed. 
In the quark sector, this pattern may also be used to check if it agrees 
with data, in analogy to what has been done \cite{mst06} in a specific 
application of $A_4$.

In conclusion, the family symmetry $\Delta(27)$ has been discussed in a 
simple model as the origin of the observed mixing pattern of neutrinos.
It is able to describe present data and has a specific prediction of 
the effective neutrino mass in neutrinoless double beta decay.

I thank Luis Lavoura for an important comment. This work was supported in 
part by the U.~S.~Department of Energy under Grant No. DE-FG03-94ER40837.

\newpage
\noindent {\bf Appendix} The matrices of the {\bf 3} representation 
of $\Delta(27)$ are given by
\begin{eqnarray}
C_1 &:& \pmatrix{1 & 0 & 0 \cr 0 & 1 & 0 \cr 0 & 0 & 1}, \\ 
C_2 &:& \pmatrix{\omega & 0 & 0 \cr 0 & \omega & 0 \cr 0 & 0 & \omega}, \\
C_3 &:& \pmatrix{\omega^2 & 0 & 0 \cr 0 & \omega^2 & 0 \cr 0 & 0 & \omega^2}, 
\\
C_4 &:& \pmatrix{0 & 1 & 0 \cr 0 & 0 & 1 \cr 1 & 0 & 0}, 
\pmatrix{0 & \omega & 0 \cr 0 & 0 & \omega \cr \omega & 0 & 0}, 
\pmatrix{0 & \omega^2 & 0 \cr 0 & 0 & \omega^2 \cr \omega^2 & 0 & 0}, \\
C_5 &:& \pmatrix{0 & 0 & 1 \cr 1 & 0 & 0 \cr 0 & 1 & 0}, 
\pmatrix{0 & 0 & \omega \cr \omega & 0 & 0 \cr 0 & \omega & 0}, 
\pmatrix{0 & 0 & \omega^2 \cr \omega^2 & 0 & 0 \cr 0 & \omega^2 & 0}, \\
C_6 &:& \pmatrix{1 & 0 & 0 \cr 0 & \omega & 0 \cr 0 & 0 & \omega^2},  
\pmatrix{\omega^2 & 0 & 0 \cr 0 & 1 & 0 \cr 0 & 0 & \omega}, 
\pmatrix{\omega & 0 & 0 \cr 0 & \omega^2 & 0 \cr 0 & 0 & 1}, \\
C_7 &:& \pmatrix{0 & 1 & 0 \cr 0 & 0 & \omega \cr \omega^2 & 0 & 0}, 
\pmatrix{0 & \omega^2 & 0 \cr 0 & 0 & 1 \cr \omega & 0 & 0}, 
\pmatrix{0 & \omega & 0 \cr 0 & 0 & \omega^2 \cr 1 & 0 & 0}, \\ 
C_8 &:& \pmatrix{0 & 0 & 1 \cr \omega & 0 & 0 \cr 0 & \omega^2 & 0}, 
\pmatrix{0 & 0 & \omega^2 \cr 1 & 0 & 0 \cr 0 & \omega & 0}, 
\pmatrix{0 & 0 & \omega \cr \omega^2 & 0 & 0 \cr 0 & 1 & 0}, \\ 
C_9 &:& \pmatrix{1 & 0 & 0 \cr 0 & \omega^2 & 0 \cr 0 & 0 & \omega},  
\pmatrix{\omega & 0 & 0 \cr 0 & 1 & 0 \cr 0 & 0 & \omega^2}, 
\pmatrix{\omega^2 & 0 & 0 \cr 0 & \omega & 0 \cr 0 & 0 & 1}, \\
C_{10} &:& \pmatrix{0 & 1 & 0 \cr 0 & 0 & \omega^2 \cr \omega & 0 & 0}, 
\pmatrix{0 & \omega & 0 \cr 0 & 0 & 1 \cr \omega^2 & 0 & 0}, 
\pmatrix{0 & \omega^2 & 0 \cr 0 & 0 & \omega \cr 1 & 0 & 0}, \\ 
C_{11} &:& \pmatrix{0 & 0 & 1 \cr \omega^2 & 0 & 0 \cr 0 & \omega & 0}, 
\pmatrix{0 & 0 & \omega \cr 1 & 0 & 0 \cr 0 & \omega^2 & 0}, 
\pmatrix{0 & 0 & \omega^2 \cr \omega & 0 & 0 \cr 0 & 1 & 0}. 
\end{eqnarray}

\bibliographystyle{unsrt}

\end{document}